\documentclass[aps,rmp,nofootinbib]{revtex4}

\usepackage{hyperref}
\setcitestyle{numbers,square}

\usepackage{graphicx}

\begin{document}

\title{Upper limit for superconducting transition temperature in electron --
phonon superconductors: very strong coupling}

\author{M.V. Sadovskii}

\affiliation{Institute for Electrophysics, Russian Academy of Sciences,
Ural Branch, Ekaterinburg 620016, Russia\\
E-mail: sadovski@iep.uran.ru}

\begin{abstract}

We present a brief review of some recent work on the problem of highest
achievable temperature of superconducting transition $T_c$ in electron -- 
phonon systems. The discovery of record -- breaking values of $T_c$ in quite a 
number of hydrides under high pressure was an impressive demonstration of 
capabilities of electron -- phonon mechanism of Cooper pairing. 
This lead to an increased interest on possible limitations of Eliashberg -- 
McMillan theory as the main theory of superconductivity in a system of electrons 
and phonons. We shall consider some basic conclusions following from this theory 
and present some remarks on the limit of very strong electron -- phonon
coupling.
We shall discuss possible limitations on the value of the coupling constant 
related to possible lattice and specific heat instability and conclude
that within the stable metallic phase the effective pairing constant may acquire
very large values. 
We discuss some bounds for 
$T_c$ derived in the strong coupling  limit and propose an elementary estimate 
of an upper limit for $T_c$, expressed via combination of fundamental physical 
constants. Finally we also briefly discuss some pessimistic estimates for $T_c$ 
of metallic hydrogen obtained in ``jellium'' model

\vskip 0.5cm

{\sl Dedicated to the memory of Mikhail Eremets}

\end{abstract}

\maketitle

\section {Introduction}

First ideas to enhance superconducting critical temperature $T_c$ were
introduced soon after the formulation of BCS theory. In 1964 Little \cite{Litt}
and Ginzburg \cite{Ginz} proposed an idea of an ``excitonic'' mechanism --
the replacement of phonons as a ``glue'' leading to Cooper pairing
by some other Boson -- type excitations with higher energies, thus changing the
Debye frequency $\omega_D$ in the preexponential factor of BCS expression for
$T_c$ by some $\omega_{ex}$ (with $\omega_{ex}>\omega_D$), which leads to the
increase of $T_c$ (probably in some lucky case up to room temperatures).
These ideas were further developed in many papers reviewed in the famous
book \cite{HTSC}.
Unfortunately, up to now this ``excitonic'' mechanism was never and nowhere
realized experimentally.
The discoveries of high -- temperature in
copper oxides (1986) and iron pnictides and chalcogenides (2008) were more or
less unrelated to these theoretical proposals and will not be discussed here.

The similar idea of the use of much larger values of $\omega_D$ in
the case of the usual electron -- phonon pairing mechanism was introduced later
by Ashcroft \cite{Ash1}, who proposed to study the metallic hydrogen (with
apparently larger $\omega_D$ due to a small mass of hydrogen ion) and
different hydrides \cite{Ash2}, which can be stable under extremely high pressures.
It is important to stress that all of these works used basically the standard
weak (or intermediate) coupling approximation of BCS (McMillan) theory.

These proposals were criticized in the notorious
paper by Cohen and Anderson \cite{CA}, where rather elegant arguments were given,
seemingly quite convincing, that characteristic scale of $T_c$ values due to
electron -- phonon or ``excitonic'' mechanism (based on exchange of Bose --
like excitations in metals) can be of the order of about 10 K only.
This paper was immediately seriously criticized in Refs. \cite{HTSC,DMK},
with the conclusion that in reality there are no such limitations.
However, the point of view expressed in Cohen -- Anderson paper
became popular in physics community (Anderson himself till the end of his life
adhered to the view expressed in Ref. \cite{CA}, though Cohen \cite{Coh} has
acknowledged the validity of arguments expressed in \cite{DMK}), so that at the time of discovery of
high -- temperature superconductivity in cuprates (1986 -- 1987), the common
belief was that the ``usual'' electron -- phonon mechanism does not allow values of
$T_c$ higher, that 30 -- 40 K. Because of this after the discovery of superconductivity
in cuprates the ``great race'' has started for new theoretical models and mechanisms
of superconductivity, which may explain the high values of $T_c$.
It is most probable, that in these compounds $T_c$ is determined by some kind
of non -- phonon pairing mechanism (e.g. due to antiferromagnetic fluctuations).
Thus the problems of superconductivity in cuprates (as well as in iron
pnictides and chalcogenides) are outside the scope of this work, which will
discuss only the electron -- phonon pairing.

The remarkable discovery by Mikhail Eremets group of superconductivity in
H$_3$S with $T_c\sim$200 K and further rapid development of experimental
studies of high -- temperature superconductivity in different hydrides
\cite{Er2,Er3,Er4,Troy} (to quote only some of the review papers) has opened
the new path to almost room -- temperatures superconductivity (though at
extremely high (megabar) pressures) and stimulated an active theoretical work
\cite{GK,Pick}. There was no doubt from the very beginning, that high -- $T_c$
superconductivity in hydrides is due to the usual electron -- phonon coupling.
So from theoretical point of view  probably the most important result of
discovery of record $T_c$ values in hydrides under high pressures, in our
opinion, is the final (and experimental!) rebuttal of the point of view
expressed in Ref. \cite{CA}, explicitly demonstrating the possibility of
achieving high - $T_c$ values (of the order of 10$^2$ K) with the common
electron -- phonon mechanism.
Most pressing now becomes the question of the upper limit
of $T_c$, which can be achieved due to this pairing mechanism.
Below we shall try to discuss this problem once again within the standard
approach, based on Eliashberg -- McMillan equations, as most successful theory, describing
superconductivity in the system of electrons and phonons in metals.

\section {Electron -- phonon interaction and Eliashberg --
McMillan theory: Strong coupling limit}

\subsection {Some general expressions and definitions}

Fr\"ohlich Hamiltonian which is commonly used to describe electron -- phonon
interaction is written as:
\begin{eqnarray}
H=\sum_{\bf p}\varepsilon_{\bf p}a^+_{\bf p}a_{\bf p}
+\sum_{\bf k}\Omega_{0{\bf k}}b^+_{\bf k}b_{\bf k}
+\frac{1}{\sqrt N}\sum_{\bf pk}g_{\bf k}a^+_{\bf p+k}
a_{\bf p}(b_{\bf k}+b^+_{-\bf k})
\label{H_Frohlih}
\end{eqnarray}
where $\varepsilon_{\bf p}$ is the conduction electron energy (counted from the
Fermi level), $\Omega_{0{\bf k}}$ is the ``bare'' phonon spectrum
{\em in the absence} of electron -- phonon interaction (which is actually
rather poorly defined in the case of a rel metal),
and we have introduced the standard notations for creation $a^+_{\bf p}$ and
annihilation $a_{\bf p}$ operators of electrons and phonons --  $b^+_{\bf k}$ and
$b_{\bf k}$,  $N$ is the number of atoms in crystal.

The matrix element of electron -- phonon interaction is usually written as:
\begin{eqnarray}
g_{\bf k}=-\frac{1}{\sqrt{2M\Omega_{0{\bf k}}}}\langle{\bf p}|
{\bf e(\bf q)}\nabla V_{ei}({\bf r})|{\bf p+q}\rangle
\equiv
-\frac{1}{\sqrt{2M\Omega_{0{\bf k}}}}I({\bf k})
\label{Fr_const}
\end{eqnarray}
where $V_{ei}$ is electron -- ion interaction potential, $M$ is the ion mass,
and ${\bf e(q)}$ is polarization vector of a phonon with frequency
$\Omega_{0\bf q}$.

McMillan \cite{McM} has derived a simple, but very general, expression for the
dimensionless electron -- phonon coupling in Eliashberg theory.
The so called Eliashberg -- McMillan function is {\em defined} \cite{McM,Sad1}
as:
\begin{eqnarray}
&&
\alpha^2(\omega)F(\omega)=
\frac{1}{N(0)}
\sum_{\bf p}\sum_{\bf p'}
|g_{\bf pp'}|^2\delta(\omega-\Omega_{\bf p-p'})\delta(\varepsilon_{\bf p})
\delta(\varepsilon_{\bf p'})=\nonumber\\
&&=\frac{1}{N(0)}\sum_{\bf p}\sum_{\bf p'}
\frac{1}{2M\Omega_{\bf p-p'}}|I({\bf p-p'})|^2\delta(\omega-\Omega_{\bf p-p'})
\delta(\varepsilon_{\bf p})
\delta(\varepsilon_{\bf p'})
\label{Elias_McMil}
\end{eqnarray}
where $\Omega_{\bf p-p'}$ is the phonon frequency of phonons and
$F(\omega)=\sum_{\bf q}\delta(\omega-\Omega_{\bf q})$ is the phonon density
of states.
As phonons typically scatter electrons in metals only in some narrow
region close to the Fermi surface we introduce the matrix element of the
gradient of electron --ion potential averaged over Fermi surface:
\begin{eqnarray}
\langle I^2\rangle=\frac{1}{[N(0)]^2}\sum_{\bf p}\sum_{\bf p'}
\left|I({\bf p-p'})\right|^2\delta(\varepsilon_{\bf p})
\delta(\varepsilon_{\bf p'})
=\frac{1}{[N(0)]^2}\sum_{\bf p}\sum_{\bf p'}
\left|\langle{\bf p}|
\nabla V_{ei}({\bf r})|{\bf p'}\rangle\right|^2)\delta(\varepsilon_{\bf p})
\delta(\varepsilon_{\bf p'})=\nonumber\\
=\langle |\langle {\bf p}|\nabla V_{ei}({\bf r})|{\bf p'}\rangle|^2\rangle_{FS}
\label{grV2}
\end{eqnarray}
Then we immediately get:
\begin{equation}
\int_{0}^{\infty}d\omega\alpha^2(\omega)F(\omega)\omega=\frac{N(0)
\langle I^2\rangle}{2M}
\label{I2M}
\end{equation}
Dimensionless electron -- phonon coupling constant is expressed now
via this Fermi -- surface average as \cite{McM,Sad1}:
\begin{eqnarray}
\lambda=2\int_{0}^{\infty}\frac{d\omega}{\omega}\alpha^2(\omega)F(\omega)
=\frac{2}{\langle\Omega^2\rangle}\int_{0}^{\infty}d\omega\alpha^2(\omega)F(\omega)\omega
\label{lambda_Eliashb_Mc}
\end{eqnarray}
where the mean square phonon frequency is defined as:
\begin{eqnarray}
\langle\Omega^2\rangle =
\frac{\int_{0}^{\infty}d\omega\alpha^2(\omega)F(\omega)\omega}
{\int_{0}^{\infty}\frac{d\omega}{\omega}\alpha^2(\omega)F(\omega)}
=\frac{2}{\lambda}\int_{0}^{\infty}d\omega
\alpha^2(\omega)F(\omega)\omega\nonumber\\
\label{aver_sq_w}
\end{eqnarray}
From this expression we can immediately see that:
\begin{equation}
\lambda=\frac{N(0)\langle I^2\rangle}{M\langle\Omega^2\rangle}
\label{McM_form}
\end{equation}
This expression gives very useful representation for $\lambda$, which is
often used in the literature and in practical calculations.

Migdal's theorem \cite{Mig} allows us to neglect vertex corrections in all
calculations of Feynman diagrams related to electron -- phonon interaction
in typical metals. The actual small parameter of perturbation theory is
$\lambda\frac{\Omega_0}{E_F}\ll 1$, where
$\lambda$ is the dimensionless constant of electron -- phonon
interaction and $\Omega_0$ is characteristic phonon frequency (e.g. of the
order of Debye frequency $\omega_D$), while $E_F$ is the Fermi energy of
electrons, which in typical metals is of the order of conduction band width.
In particular this leads to a common
belief, that vertex corrections in this theory can be neglected even in case of
$\lambda > 1$ (up to the values of $\lambda\sim\frac{E_F}{\Omega_0}\gg 1$),
as inequality $\frac{\Omega_0}{E_F}\ll 1$ holds in typical metals.
This fact is the cornerstone of Eliashberg -- McMillan theory for superconductors
which allows the description of the so called strong coupling superconductivity
outside the usual weak coupling limit of BCS theory
\cite{El1,El2,McM,AD,AM,CAEK,Sad1}.

\subsection {{\em Lower} bound for $T_c$ in Eliashberg -- McMillan theory
and strong coupling limit}

Limitations on the value of $T_c$ in Eliashberg -- McMillan theory in the
limit of very strong coupling can be derived analytically \cite{AD,Sad1}.
In the following we shall not consider the role of direct Coulomb repulsion of
electrons within the Cooper pair, which is accounted for in the complete
Eliashberg -- McMillan theory, limiting ourselves only to electron -- phonon
interaction. The accounting for Coulomb contributions is not especially difficult
\cite{McM} and reduces at the end to introduction of the usual Coulomb
pseudopotential $\mu^{\star}$, which in typical metals is rather small and not
so important in the limit of very strong coupling with phonons,
which will be of the main interest for us in the following.

From the general system of Eliashberg -- McMillan equations directly follows
the linearized  equation for the gap $\Delta(\omega_n)$ \cite{Sad1},
determining $T_c$:
\begin{eqnarray}
\Delta(\omega_n)Z(\omega_n)=\pi T\sum_{n'}\int_{0}^{\infty}\alpha^2(\omega)
F(\omega)
D(\omega_n-\omega_{n'};\omega)\frac{\Delta(\omega_{n'})}{|\omega_{n'}|}
\label{lin_delta_gen}
\end{eqnarray}
where $\omega_n=(2n+1)\pi T$ are usual Matsubara frequencies of electrons.
The renormalization factor  $Z(\omega_n)$ is determined from:
\begin{eqnarray}
&&1-Z(\omega_n)=\frac{\pi T}{\omega_n}\sum_{n'}\int_{-\infty}^{\infty}d\xi\int_{0}^{\infty}d\omega
\alpha^2(\omega)F(\omega)
D(\omega_n-\omega_{n'};\omega)
\frac{\omega_{n'}}
{|\omega{_n'}|}
\label{lin_Zgen}
\end{eqnarray}
and phonon Green's function is expressed as:
\begin{equation}
D(\omega_n-\omega_{n'};\omega)=\frac{2\omega}{(\omega_n-\omega_{n'})^2+\omega^2}
\label{Dw}
\end{equation}
Actually, Eq. (\ref{lin_delta_gen}) represent the system of linear equations
for $\Delta(\omega_n)$. Let us consider here first the
term with $n=0$. Then, leaving in the sum in Eq. (\ref{lin_Zgen}) only
the contribution from $n'=0$, we obtain:
\begin{equation}
Z(0)=1+\lambda
\label{Zlamb}
\end{equation}
which represents the usual electron mass -- renormaliztion factor due to
electron -- phonon interaction: $m^{\star}=m(1+\lambda)$. Its substitution into
Eq. (\ref{lin_delta_gen}) for  $n=0$ just
cancels the similar (corresponding to $n'=0$) term in the r.h.s.,
so that the equation for $\Delta(0)=\Delta(\pi T)$ takes the form:
\begin{equation}
\Delta(0)=\pi T\sum_{n'\neq 0}\int_{0}^{\infty}\alpha^2(\omega)
F(\omega)\frac{2\omega}{(\pi T-\omega_{n'})^2+\omega^2}\frac{\Delta(\omega_{n'})}{|\omega_{n'}|}
\label{lin_delta_0}
\end{equation}
All terms in the r.h.s. here are positive. Let us leave only the contribution
from $n'=-1$, then after simple algebra with the account of
$\Delta(0)=\Delta(\pi T)=\Delta(-\pi T)=\Delta(-1)$ we immediately obtain
the {\em inequality} \cite{AD,Sad1}:
\begin{equation}
1>\int_{0}^{\infty}d\omega\frac{2\alpha^2(\omega)F(\omega)\omega}
{(2\pi T)^2+\omega^2}
\label{AD_ineq}
\end{equation}
Putting $T=T_c$ in (\ref{AD_ineq}) we obtain the {\em lower} bound for
$T_c$. In particular, in the model with Einstein spectrum of phonons
$F(\omega)=\delta(\omega-\Omega_0)$ and this inequality is immediately rewritten as:
\begin{equation}
1>2\alpha^2(\Omega_0)\frac{\Omega_0}{(2\pi T)^2+\Omega_0^2}=
\lambda\frac{\Omega^2_0}{(2\pi T)^2+\Omega_0^2}
\label{AD_in}
\end{equation}
so that for $T_c$ we get:
\begin{equation}
T_c>\frac{1}{2\pi}\sqrt{\lambda-1}\Omega_0
\label{TcAD}
\end{equation}
which for $\lambda\gg 1$ reduces to:
\begin{equation}
T_c>\frac{1}{2\pi}\sqrt{\lambda}\Omega_0\approx 0.16\sqrt{\lambda}\Omega_0
\label{TcA_D}
\end{equation}
\begin{figure}
\includegraphics[clip=true,width=0.8\textwidth]{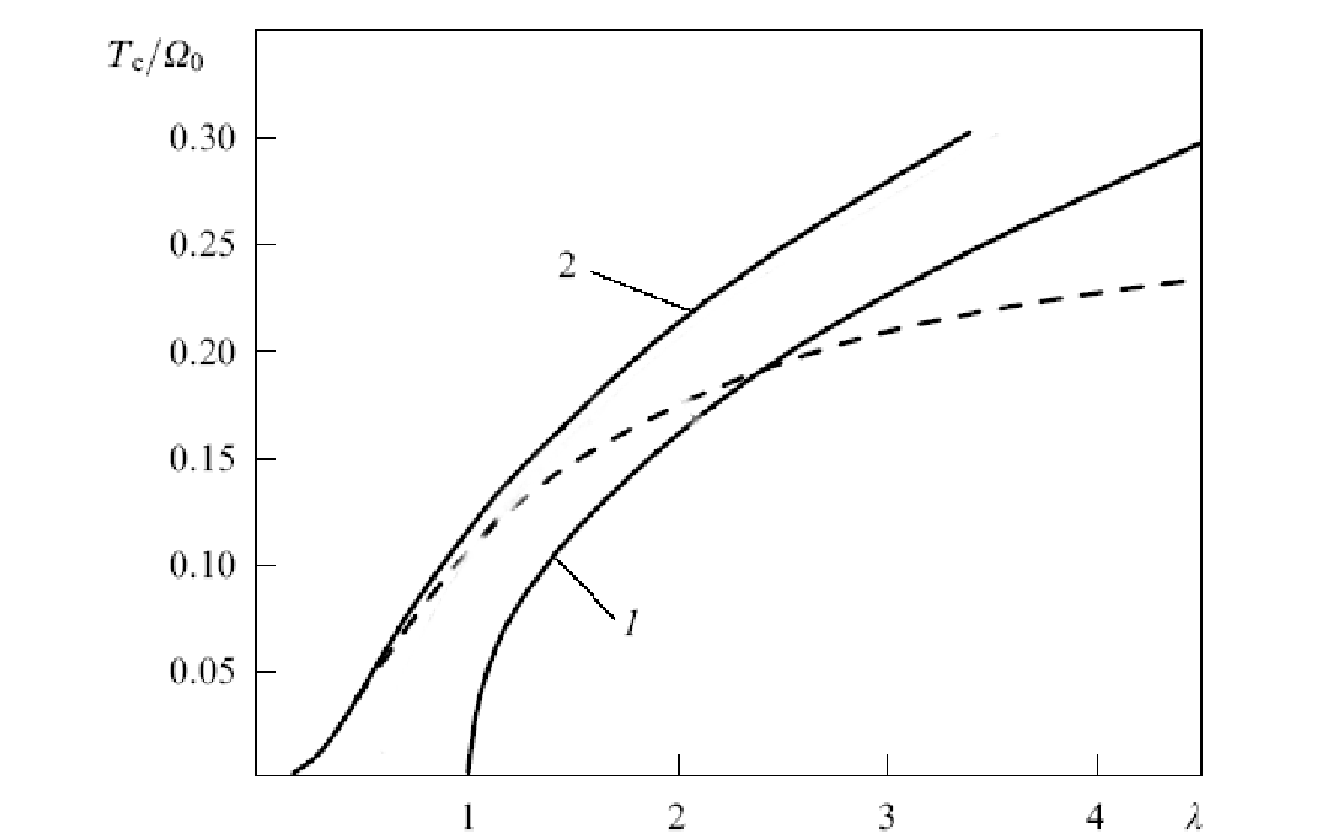}
\caption{Temperature of superconducting transition in Einstein model
of phonon spectrum in units of $T_c/\Omega_0$ as a function of pairing constant
$\lambda$ \cite{AD}: 1 -- lower bound (\ref{TcAD}),
2 -- numerically exact solution of the full system of equations \cite{AD}.
McMillan expression for $T_c$ \cite{McM} is shown by dashed line (for the case of
$\mu^{\star}$=0).}
\label{TcADcomp}
\end{figure}

For the model of phonon spectrum consisting of discrete set of Einstein
phonons:
\begin{eqnarray}
\alpha^2(\omega)F(\omega)=\sum_i\alpha^2(\Omega_i)\delta(\omega-\Omega_i)\nonumber\\
=\sum_i\frac{\lambda_i}{2}\Omega_i\delta(\omega-\Omega_i)
\label{El-Mc-discr}
\end{eqnarray}
In this case from (\ref{aver_sq_w}) we simply obtain:
\begin{equation}
\langle\Omega^2\rangle =\frac{1}{\lambda}\sum_i\lambda_i\Omega_i^2
\label{sqfrqav}
\end{equation}
where $\lambda=\sum_i\lambda_i$.
In this case the inequality (\ref{AD_ineq}) reduces to:
\begin{equation}
1>\sum_i\lambda_i\frac{\Omega_i^2}{(2\pi T)^2+\Omega_i^2}
\label{ADIneqdiskr}
\end{equation}
which in the limit of very strong coupling
immediately gives the natural generalization of (\ref{TcA_D}):
\begin{equation}
T_c > \frac{1}{2\pi}\sqrt{\lambda\langle\Omega^2\rangle}
\label{Tc_Al_Dy}
\end{equation}
Numerically exact solution of the full system of Eq. (\ref{lin_delta_gen}),
performed in Ref. \cite{AD} leads to the final expression for $T_c$ in the
strong coupling limit of $\lambda\gg 1$ with replacement of $1/2\pi=$0.16
in (\ref{TcA_D}) or (\ref{Tc_Al_Dy}) by 0.182:
\begin{equation}
T_c = 0.182\sqrt{\lambda\langle\Omega^2\rangle}
\label{Tc_Al_Dyn}
\end{equation}
It is obvious, that even the simplest solution (\ref{TcAD}) is quite sufficient
for qualitative estimates of $T_c$ in the limit of very strong coupling.
The general situation is illustrated in Fig. \ref{TcADcomp}. From this figure
it can be seen, in particular, that asymptotic behavior of $T_c$
for $\lambda\gg$1 (\ref{Tc_Al_Dyn}) with coefficient 0.182,  approximates
the values of critical temperature rather well  already starting from the values
of $\lambda>$1.5-2.0. The remarkable result here
is the replacement of exponential dependence of $T_c$ on the coupling constant,
typical for the weak coupling BCS or intermediate coupling McMillan
approximations, by the square root dependence, leading to monotonous
and seemingly unlimited growth of $T_c$ with increasing $\lambda$.
Then an important question arises -- are there any limitations to the growth
of $\lambda$ ?

\section {Possible limitations for electron --
phonon coupling}

\subsection {Fr\"ohlih instability}

\begin{figure}
\includegraphics[clip=true,width=0.8\textwidth]{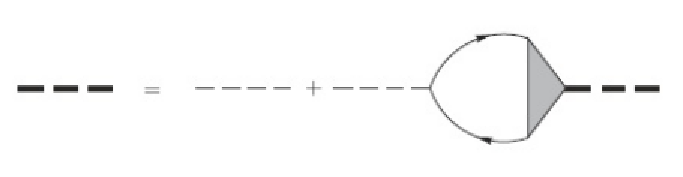}
\caption{Dyson equation for the full (``dressed'') phonon Green's function.}
\label{Phon_Dress}
\end{figure}

The general expression for phonon Green's function, taking into account the interaction
with electrons, is given by Dyson equation shown in Fig. {\ref{Phon_Dress}}.
Then such  ``dressed'' phonon Green's function can be written as:
\begin{equation}
D({\bf k}\omega)=\frac{2\Omega_{0{\bf k}}}
{\omega^2-\Omega^2_{\bf k}+i\delta}
\label{fonGrdress}
\end{equation}
where the renormalized phonon spectrum is determined from the
equation:
\begin{equation}
\Omega^2_{\bf k}=\Omega_{0{\bf k}}^2\left[1+\frac{2|g_{\bf k}|^2}{\Omega_{0{\bf k}}}
\Pi({\bf k},\Omega_{\bf k})\right]
\label{ph_spectr}
\end{equation}
In adiabatic approximation, taking into account Migdal theorem, polarization operator
here can be taken as a simple loop.
With rather high accuracy in polarization operator here we can put $\omega=0$,
${\bf k=0}$  so that it reduces to $\Pi(0,0)=-2N(0)$ \cite{Sad1}.
Then the phonon spectrum, renormalized by interaction with
electrons, is determined by:
\begin{equation}
\Omega^2_{\bf k}=\Omega_{0{\bf k}}^2\left[1+\frac{2|g_{\bf k}|^2}{\Omega_{0\bf k}}
\Pi(0,0)\right]=\Omega_{0\bf k}^2\left[1-2\lambda_0^k\right]
\label{ph_spectr_inst}
\end{equation}
where we have introduced the usual definition of dimensionless coupling constant of
electron -- phonon interaction \cite{Sad1}:
\begin{equation}
\lambda_0^k=\frac{2|g_{\bf k}|^2 N(0)}{\Omega_{0{\bf k}}}
\label{dimepp}
\end{equation}
In this (rough enough) approximation the relatively small damping of phonons due
to electron -- phonon interaction is just neglected. It can be taken into
account by more accurate treatment of the imaginary part of polarization
operator which is just zero in the approximation used here.

The spectrum given by Eq. (\ref{ph_spectr_inst}) signifies the lattice instability
for $\lambda_0^k>1/2$, when the square of the phonon frequency becomes negative.
This instability was correctly interpreted already in the early paper by
Fr\"ohlich \cite{Fro}, where it was observed for the first time.
Let us rewrite the ``dressed'' Green's function (\ref{fonGrdress}) identically
as:
\begin{equation}
D({\bf k}\omega)=\frac{2\Omega_{\bf k}}
{\omega^2-\Omega^2_{\bf k}+i\delta}\frac{\Omega_{0{\bf k}}}{\Omega_{\bf k}}
\label{fonGrdressed}
\end{equation}
Then it becomes clear that during diagram calculations
using from the very beginning this renormalized
Green's function of phonons, the physical coupling constant of electron --
phonon coupling takes the form (instead of (\ref{dimepp})):
\begin{equation}
\lambda^k=\frac{2|g_{\bf k}|^2 N(0)}{\Omega_{\bf{k}}}
\frac{\Omega_{0{\bf k}}}{\Omega_{\bf k}}=
\frac{2|g_{\bf k}|^2 N(0)}{\Omega_{0{\bf k}}}\frac{\Omega^2_{0{\bf k}}}{\Omega^2_{\bf k}}
=\lambda_0^k\frac{\Omega^2_{0{\bf k}}}{\Omega^2_{\bf k}}
\label{dimepsc}
\end{equation}
or, using (\ref{ph_spectr_inst}):
\begin{equation}
\lambda^k=\frac{\lambda_0^k}{1-2\lambda_0^k}
\label{dimepscons}
\end{equation}
We see, that for $\lambda_0^k\to 1/2$ the {\em renormalized} coupling constant
$\lambda^k$ monotonously grows and finally diverges.
It is this constant that determines the ``true''
value of electron -- phonon interaction (with ``dressed'' phonons) and there
are  no limitations for its value at all. This physical picture was discussed
in detail in already in Ref. \cite{HTSC}.

In the model with single Einstein phonon, which is a reasonable approximation for an optical
phonon, we have $\Omega_{\bf k}=\Omega_0$ and we can forget about dependence of
the coupling constant on phonon momentum, so that:
\begin{equation}
\lambda_0=\frac{2g_0^2 N(0)}{\Omega_0}
\label{dimep_ein}
\end{equation}
\begin{equation}
\Omega^2=\Omega_0^2[1-2\lambda_0]
\label{phn_spc_ein}
\end{equation}
\begin{equation}
\lambda=\frac{2g_0^2 N(0)}{\Omega_0}
\left(\frac{\Omega_0}{\Omega}\right)^2=\frac{\lambda_0}{1-2\lambda_0}
\label{dimepsc_ein}
\end{equation}
Equation (\ref{dimepscons}) can be reversed and we can write:
\begin{equation}
\lambda_0^{k}=\frac{\lambda^k}{1+2\lambda^k}
\label{gbare}
\end{equation}
expressing nonphysical ``bare'' constant of electron -- phonon coupling
$\lambda_0^k$ via the ``true'' physical coupling constant $\lambda^k$.
Using this relation in the equation for renormalized phonon spectrum
(\ref{ph_spectr_inst}), we can write it as:
\begin{equation}
\Omega^{2}_{\bf k}=\Omega_{0{\bf k}}^2\left[1-\frac{2\lambda^k}{1+2\lambda^k}\right]
=\Omega_{0{\bf k}}^2\frac{1}{1+2\lambda^k}
\label{phn_spc_dressed}
\end{equation}
so that in this representation there is no instability of spectrum (lattice),
and the growth of $\lambda^k$ just leads to continuous ``softening'' of
spectrum due to the growth of electron -- phonon coupling.

In a model of Einstein phonon all relations simplify and we get:
\begin{equation}
\lambda_0=\frac{\lambda}{1+2\lambda}
\label{gbare_eins}
\end{equation}
\begin{equation}
\Omega=\frac{\Omega_0}{\sqrt{1+2\lambda}}
\label{phn_spc_dressed_eins}
\end{equation}

In Eliashberg -- McMillan formalism, where we perform the averaging over the momenta of
electrons over the Fermi surface, Eliashberg --  McMillan function
$\alpha^2(\omega)F(\omega)$, naturally should be determined bu the physical
(renormalized) spectrum of phonons.
In particular case of Einstein phonon it immediately reduces to
(\ref{dimepsc_ein}) and there are no limitations on the value of $\lambda$.

In self -- consistent derivation of Eliashberg equations we have to use the
the phonon Green's function a ``dressed''form (\ref{fonGrdress}) or
(\ref{fonGrdressed}), which corresponds to the physical
(renormalized) phonon spectrum. In this case we {\em do not have to include}
corrections to this function due to electron -- phonon interaction, as they are
already taken into account in the renormalized phonon spectrum
(\ref{ph_spectr_inst}).

It should be noted that the value of critical coupling constant obtained above, at which
Fr\"ohlich instability of phonon spectrum appears, is obviously directly related to the use
of the simplest expression for polarization operator of the gas of free electrons,
which was calculated neglecting vertex corrections and
self -- consistent ``dressing'' of electron Green's functions entering the loop.
Naturally, even in the simplest cases like the problem with Einstein spectrum
accounting for these higher corrections, as well as more realistic structure of
electron spectrum in a lattice, can somehow change the value of  $\lambda_0$,
corresponding to instability of the ``bare'' phonon spectrum, so that it will
differ from 1/2. In this sense it is better to speak about instability at some
``critical'' value  $\lambda_0^{c}\sim 1/2$.

This analysis can be significantly improved within the simplified Holstein
model, where the electron -- phonon interaction is considered to be
local (single site), which allows solving this model
using the dynamical mean field theory (DMFT) approach \cite{GKKR},
which becomes (numerically) exact in the limit of lattice of infinite
dimensions (infinite number of nearest neighbors). Such analysis was performed
e.g. in Ref. \cite{Gunn}, using as quantum Monte -- Carlo (QMC) as impurity
solver of DMFT.
The usual behavior of Fr\"ohlich theory is nicely reproduced with slightly
changed $\lambda_{0}^{c}$=0.464.
Behavior similar to Eq. (\ref{phn_spc_dressed_eins}) was obtained also for
the renormalized phonon frequency $\Omega$.,

The instability appearing at $\lambda_0=\lambda_{0}^{c}$ in Holstein model
with half -- filled bare band, was convincingly
interpreted in Ref. \cite{Bull} as transition into the state of
{\em bipolaron insulator}. Until this transition the system remains metallic
and is nicely described by Eliashberg theory (with insignificant numerical
corrections).

It should be stressed here, that all conclusions on instability of metallic
phase were done above in the framework of purely {\em model} approach
and in terms of ``bare'' parameters of $\lambda_0$ and $\Omega_0$,
which, as was often noted in the literature, are not so well defined physically.
The problem here is that the phonon spectrum in a
metal, considered as system of ions and electrons, is usually
calculated in adiabatic approximation \cite{BK}. This spectrum is relatively
weakly renormalized due to nonadiabatic effects, which are small over the
parameter $\sqrt{\frac{m}{M}}$ \cite{BK,Geilik}.
In this respect, it is drastically different from the ``bare'' spectra of
Fr\"ohlich or Holstein models, which, as we have seen above, is significantly
renormalized by electron -- phonon interaction. The physical meaning of the
``bare'' spectrum $\Omega_0$ in these models remains not so clear, in contrast
to phonon spectrum in metals, calculated in adiabatic approximation.

\subsection {Specific heat instability?}

Recently Semenok  et al. \cite{AY}  proposed a possible limitation for  $\lambda$
due a certain electronic specific heat instability previously derived
in Yuzbashyan and Altshuler \cite{YA} within Eliashberg -- McMillan theory .
It was claimed that electronic specific heat in Eliashberg -- McMillan theory
becomes negative for the values of $\lambda>\lambda_{\star}$ in a certain
temperature interval, signifying thermodynamic instability of electron -- phonon
system. For Einstein model it was shown that $\lambda_{\star}=$3.69 (for Debye
model of phonon spectrum $\lambda_{\star}=$4.72). The use of this value in
Allen -- Dynes expression for $T_c$ (\ref{Tc_Al_Dyn}), immediately leads to
$T_c\approx 0.35\Omega$ as an upper limit for superconducting transition
temperature (for Einstein model).

However the authors of Ref. \cite{SBC} disagreed and explicitly shown that
the total specific heat in Eliashberg -- McMillan theory remains positive
for all parameters of the model, at least until adiabatic approximation is
valid. i.e. $\lambda < \frac{E_F}{\Omega_0}$. The essence of argumentation is
as follows. The free energy per unit volume in Eliashberg -- McMillan theory
can be shown \cite{SBC} to be determined as:
\begin{equation}
F=F_{free}+\frac{T}{2}\sum_{m\bf k}\log[-D^{-1}({\bf k},i\omega_m)]
\label{Free}
\end{equation}
where $F_{free}$ is the free energy of free electrons and
$D({\bf k},i\omega_m)$ is the renormalized phonon Green's function
(\ref{fonGrdress}), (\ref{fonGrdressed}) in Matsubara representation,
calculated in the simplest approximation used above. Actually electron --
phonon interaction enters here only through this ``dressed'' function.

After the detailed calculations an explicit expressions for $F$ and specific
heat $C_{ep}(T)=-Td^2F/d^2T$ can be derived \cite{SBC} of which we quote only
the limiting forms for $C_{ep}(T)$ for the case of Einstein phonons, using
our notations. For $2\pi T\ll\Omega$ (with $\Omega$ given by
Eq. (\ref{phn_spc_ein})):
\begin{equation}
C_{ep}=\frac{2\pi^2}{3}N(0)T\left(1+\frac{\lambda_0}{1-2\lambda_0}\right)
=\frac{2\pi^2}{3}N(0)T(1+\lambda)
\label{C_lowT}
\end{equation}
which is the standard result for low -- temperature
contribution to specific heat with the account of electronic mass renormalization
due to interaction with phonons: $m^{\star}=m(1+\lambda)$.
For high temperatures $2\pi T\gg\Omega$:
\begin{equation}
C_{ep}=\frac{2\pi^2}{3}N(0)\left(T+\frac{6E_F}{\pi^2}-\frac{3}{\pi^2}\lambda_0
\frac{\Omega_0^2}{T}\right)
\label{C_highT}
\end{equation}
Consider now this last expression. The first term here is just the usual
contribution from free electrons, the second is the contribution from free
phonons with effective frequency, {\em renormalized} by the interaction with electrons,
and the third term is the contribution from electron -- phonon interaction.
Without the middle term, the specific heat becomes negative below
$T\approx 0.39\sqrt{\lambda_0}\Omega_0=0.39\sqrt{\lambda}\Omega$ which is
higher that superconducting transition temperature in the strong coupling limit.
However, with the account of the middle term the full $C_{ep}(T)$ is never
negative.

The authors of Ref. \cite{YA} argued that the negative contribution from the
third term in (\ref{C_highT}) indicates the normal state instability below a
certain $T$, despite the total $C_{ep}(T)$ remains positive, as the contribution
from free phonons has nothing to do with electrons. However, both positive and
negative contributions here come from the free energy term
$\frac{T}{2}\sum_{m\bf k}\log[-D^{-1}({\bf k},i\omega_m)]$, so that both terms
should be treated equally. Thus, the electron -- phonon coupling generates the
negative contribution to $C_{ep}(T)$ and simultaneously gives the rise to
{\em much larger} positive $T$ -- independent contribution. The detailed
numerical calculations of specific heat for all temperatures and different
values of parameters of the model, performed in Ref. \cite{SBC}, has shown that
specific heat remains always positive. In our opinion this clearly shows, that
there is no specific heat instability discussed in Refs. \cite{AY,YA}, at least
within the standard formalism of equilibrium statistical mechanics.

However, the situation may be more complicated. It was shown in Ref. \cite{AYP}
using the standard kinetic equation approach, that thermal equilibrium between
electrons and phonons can actually become unstable for large values of electron --
phonon coupling constant. It was claimed that the negative values of electronic
specific heat only are sufficient for such instability, leading to the difference
between the temperatures of electrons and phonons. These results stress the
importance of further studies of possible specific heat instabily in electron --
phonon system at large couplings.

\section {{\em Upper} bound for $T_c$ in the very strong coupling limit}

As we have seen above in the limit of very strong coupling $\lambda\gg 1$
solution of Eliashberg -- McMillan equations gives the following expression
for $T_c$:
\begin{equation}
T_c= 0.18\sqrt{\lambda\langle\Omega^2\rangle}
\label{AD_asymp}
\end{equation}
It may seem now, that there is no limit for $T_c$ growth due to electron --
phonon pairing mechanism in the limit of very strong coupling.
The only more or less obvious limit is
related to the limits of adiabatic approximation, which is usually considered
to be the cornerstone of Eliashberg theory.

In the model with Einstein spectrum of phonons we simply have:
$\langle\Omega^{2}\rangle^{1/2}=\Omega$,
where $\Omega$ is assumed to be the renormalized phonon frequency.
Then (\ref{AD_asymp}) reduces to:
\begin{equation}
T_c=0.18\sqrt{\lambda}\Omega
\label{ADyn}
\end{equation}
so that seemingly for $\lambda\gg 1$ we can, in principle, obtain even
$T_c>\Omega$.
However, if we remember the renormalization of phonon spectrum and take into
account Eq. (\ref{phn_spc_dressed_eins}), we immediately obtain from
Eq. (\ref{ADyn}):
\begin{equation}
T_c=0.18\sqrt{\lambda}\Omega=0.18\Omega_0\sqrt\frac{\lambda}{1+2\lambda}
\label{ADynes}
\end{equation}
which in the limit of $\lambda\gg 1$ gives
\begin{equation}
T_{c}^{max}\approx 0.13\Omega_0,
\label{Tcmax}
\end{equation}
because of significant softening of phonon spectrum.
At the same time, as noted above, the physical meaning of ``bare'' frequency
$\Omega_0$ in a metal is rather poorly defined, and in particular it can not
be determined from any experiments.

If we just forget about ``bare'' spectrum of phonons and consider parameters
$\Omega$ and $\lambda$ {\em independent}, we can obtain from Eq. (\ref{ADyn})
very high values of $T_c$. A certain , though rather artificial model, leading
precisely to this kind of behavior was recently introduced in
Ref. \cite{KivBerg}. It considered the interaction of $N$--component electrons
with  $N\times N$--component system of Einstein phonons in the limit of
$N\to\infty$. It was shown that in this model the renormalization of phonon
spectrum due to interaction with conduction electrons is suppressed, so that
in the limit of very strong coupling with 1$\ll\lambda\ll N$ we always obtain
Allen -- Dynes estimate (\ref{ADyn}) with $\Omega=\Omega_0$.

However, the problem here is, that in real situation we never can consider
$\Omega$ and $\lambda$ as independent parameters simply because of the general
relations (\ref{lambda_Eliashb_Mc}) and (\ref{aver_sq_w}), which express
$\lambda$ and $\langle\Omega^2\rangle$ via integrals of Eliashberg -- McMillan
function $\alpha^2(\omega)F(\omega)$. In fact, we may rewrite the expression
for $T_c$ in the region of very strong coupling as:
\begin{equation}
T_c=0.18\sqrt{\lambda\langle\Omega^2\rangle}=0.25\left[\int_{0}^{\infty}d\omega
\alpha^2(\omega)F(\omega)\omega\right]^{1/2}
\label{TcaFw}
\end{equation}
We see, that this expression for $T_c$ is completely determined
by the integral of $\alpha^2(\omega)F(\omega)$ over the phonon spectrum, while
there is no explicit dependence of $T_c$ on $\lambda$ and
$\langle\Omega^2\rangle$ separately.

Experimental discovery of high -- temperature superconductivity in hydrides
under high (megabar) pressures \cite{Er1} stimulated the search for the ways
to achieve superconductivity at room temperature \cite{Er2,Er3,Er4,Troy}.
At the moment the common view \cite{GK,Pick} is  that the high -- temperature
superconductivity in hydrides can be described in the framework of the standard
Eliashberg -- McMillan theory. Within this theory
many attempts were undertaken to estimate the maximal achievable
superconducting transition temperature and the discussion of some of these
attempts can be found in the reviews \cite{GK,Pick,Sad1}.

In Ref. \cite{Leav} a another simple inequality for $T_c$ was proposed,
limiting its value by the square $A$ under $\alpha^2(\omega)F(\omega)$:
\begin{equation}
T_c\leq 0.2309\int_{0}^{\infty}d\omega\alpha^2(\omega)F(\omega)\equiv
0.2309 A
\label{Leavns}
\end{equation}
For the case of Einstein spectrum of phonons this can be
rewritten as:
\begin{equation}
T_c\leq 0.115\lambda\Omega_0
\label{Leavs}
\end{equation}
This inequality is relatively often used in the literature.

\begin{figure}
\includegraphics[clip=true,width=0.8\textwidth]{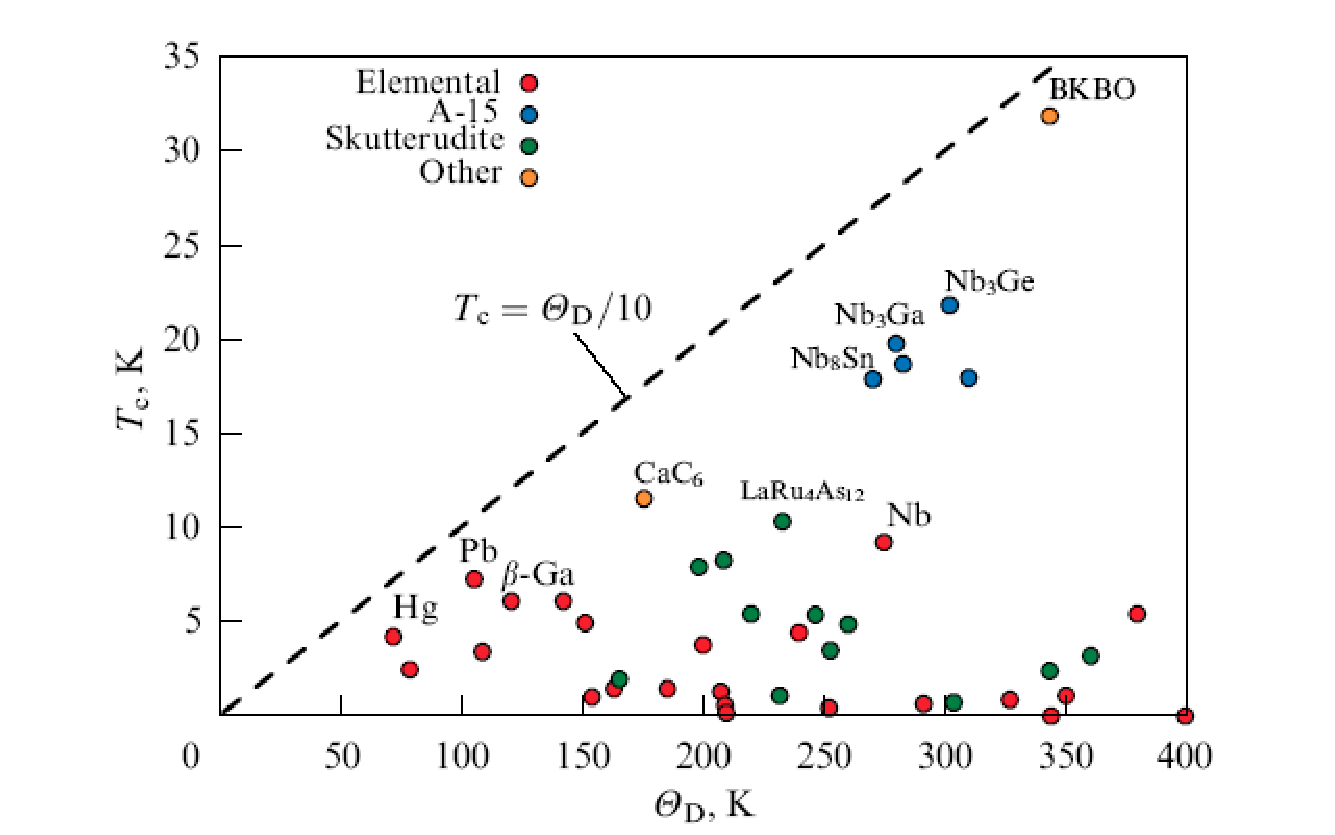}
\caption{Experimental values of the temperature of superconducting transition
for conventional superconductors dependence on their Debye temperature
$\Theta_D$
\cite{EKS}.}
\label{MaxTcKiv}
\end{figure}

In Ref. \cite{EKS} a semi empirical limit for $T_c$ was proposed for conventional
semiconductors, which can be written in a very simple form:
\begin{equation}
k_BT_c\leq A_{max}\Theta_D=A_{max}\hbar\Omega_D
\label{Kiv_Scal}
\end{equation}
where $A_{max}\approx$ 0.10, and $\Theta_D=\hbar\Omega_D$ is Debye temperature,
which may be determined e.g. from standard measurements of specific heat.
This inequality obviously correlates with $T^{max}_{c}$, obtained above in the
limit of $\lambda\to\infty$ in Eq. (\ref{ADynes}), if we identify $\Omega_0$
with $\Omega_D$.
It is seen from Fig. \ref{MaxTcKiv} this limitation is satisfied for most of
conventional superconductors \cite{EKS}.

In the recent paper \cite{Tr} a new upper limit for $T_c$ was proposed,
expressed via certain combination if fundamental constants. Below we  show
that with minor modifications such $T_c$ limit follows directly from
Eliashberg -- McMillan theory.
The relation of $\lambda$ and $\langle\Omega^2\rangle$ is clearly expressed by
McMillan's formula for $\lambda$ (\ref{McM_form}).
Equation (\ref{McM_form}) gives very
useful representation for the coupling constant $\lambda$, which is
routinely used in the literature and in practical (first -- principles)
calculations \cite{Pick,WPick}.

Using Eq. (\ref{McM_form}) in Eq. (\ref{AD_asymp}) we immediately
obtain:
\begin{equation}
T_c^{\star}=0.18\sqrt{\frac{N(0)\langle I^2\rangle}{M}}
\label{AD_McMl}
\end{equation}
where $\langle I^2\rangle$ was defined in Eq. (\ref{grV2}).
Both $\lambda$ and $\langle\Omega^2\rangle$ just drop out from the expression
for $T_c^{\star}$, which is now expressed via Fermi surface averaged matrix
element of the gradient of electron -- ion potential, ion mass and electronic
density of states at the Fermi level.

As was already noted, all parameters entering this expression can be rather
simply obtained during the first -- principles calculations of $T_c$  for
specific materials (compounds) \cite{Pick,WPick}. Let us also stress that the value
of $T_c^{\star}$ defined in Eq. (\ref{AD_McMl}), calculated for any specific
material does not have any direct relation to real value of $T_c$, but just
defines precisely the upper limit of $T_c$, which ``would be achieved'' in the
limit of strong enough electron -- phonon coupling.
Below we shall present some rough heuristic (dimensional) estimates of its
value \cite{Sad2}.

In the following we assume to be dealing with three -- dimensional metal
with cubic symmetry with an elementary cell with lattice constant
$a$ and just one conduction electron per atom. Then we can estimate the
density of states at the Fermi level as for free electrons:
$N(0)=\frac{mp_F}{2\pi^2\hbar^3}a^3$,
where $p_ F\sim\hbar/a$ is the Fermi momentum, $m$ is the mass of free (band)
electron. Electron -- ion potential (single -- charged ion, $e$ is electron
charge) can be estimated as:
\begin{equation}
V_{ei}\sim\frac{e^2}{a}\sim e^2p_F/\hbar
\label{Vei}
\end{equation}
so that its gradient is:
\begin{equation}
\nabla V_{ei}\sim\frac{e^2}{a^2}\sim e^2p_F^2/\hbar^2
\label{gradV}
\end{equation}
Then we easily obtain the following estimate of (\ref{grV2}):
\begin{equation}
I^2\sim \left(\frac{e^2}{a^2}\right)^2\sim (e^2p_F^2/\hbar^2)^2
\label{mxel}
\end{equation}
which is probably too optimistic, as we neglected all the fine details,
which were analyzed e.g. in \cite{WPick}. Here we also dropped different
numerical factors of the order of unity, which
are obviously not so important for our  order of magnitude estimates.
Now we obtain an estimate for $T_c^{\star}$ from Eq. (\ref{AD_McMl}) as:
\begin{equation}
T_c^{\star}\sim 0.2\sqrt{\frac{m}{M}}\frac{e^2}{\hbar v_F}E_F
\label{Tstar}
\end{equation}
where $E_F=p_F^2/2m$ is Fermi energy, $v_F=p_F/m$ is electron velocity at the
Fermi surface. The value of $\frac{e^2}{\hbar v_F}$, as is well known,
represents the dimensionless coupling for Coulomb interaction and for
typical metals it is of the order of or greater than unity. The factor of
$\sqrt{\frac{m}{M}}$ determines isotopic effect.

Let us measure length in units of Bohr radius $a_B$ introducing the standard
dimensionless parameter $r_s$ by the relation $a^3=\frac{4\pi}{3}(r_sa_B)^3$.
Then we have:
\begin{equation}
a\sim r_sa_B=r_s\frac{\hbar^2}{me^2}=r_s\frac{\hbar}{mc\alpha}
\label{a_b}
\end{equation}
where we have introduced the fine structure constant
$\alpha=\frac{e^2}{\hbar c}$. Correspondingly the Fermi momentum is given by:
\begin{equation}
p_F\sim\frac{\hbar}{r_sa_B}=\frac{me^2}{\hbar r_s}=\frac{mc}{\hbar r_s}\alpha
\label{PFer}
\end{equation}
Then $T_c^{\star}$ (\ref{AD_McMl}) can be rewritten as:
\begin{equation}
T_c^{\star}\sim\frac{0.2}{r_s}\sqrt{\frac{m}{M}}\alpha^2mc^2/2=
\frac{0.2}{r_s}\sqrt\frac{m}{M}\frac{me^4}{2\hbar^2}=
\frac{0.2}{r_s}\sqrt\frac{m}{M}Ry
\label{Tc_star}
\end{equation}
where $Ry=me^4/2\hbar^2\approx$ 13.6 eV is the Rydberg constant.
Here we have obtained the same combination of fundamental (atomic) constants,
which was actually suggested in Ref. \cite{Tr}, by some quite different
reasoning, as determining the upper limit of superconducting critical
temperature.
However, our expression contains an extra factor of $r_s^{-1}$, which
necessarily reflects the specifics of a material under consideration
(density of conduction electrons), so that the value of $T_c^{\star}$ is in no
sense universal. Apparently in Ref. \cite{Tr} it was somehow implicitly
assumed the value of $r_s=1$.

For  metallic hydrogen $M$ is equal to proton mass and we have
$\sqrt\frac{m}{m_p}\sim 0.02$, so that for $r_s=1$ we get an estimate of
$T_c^{\star}\sim 650$ K. This is in nice agreement with the result of
$T_c=$ 600 K, obtained in Ref. \cite{SavMaks,EGM} solving Eliashberg equations
for FCC lattice of metallic hydrogen with $r_s=1$, taking into account the
calculated softening of the phonon spectrum, leading to realizations of very
strong coupling ($\lambda=6.1$).

\section{Metallic hydrogen and the ``jellium'' model:  insufficiency of the weak -- coupling}

In the recent paper \cite{VdM} an elegant study of superconductivity of metallic 
hydrogen was performed within the ``jellium'' model \cite{HTSC,DG}. 
In this model it is rather easy to calculate all the relevant parameters to
calculate $T_c$ (cf. section Methods of Ref. \cite{VdM}) in weak -- coupling
BCS approach or for the general case of  Eliashberg -- McMillan theory .

\begin{figure}
\includegraphics[clip=true,width=0.8\textwidth]{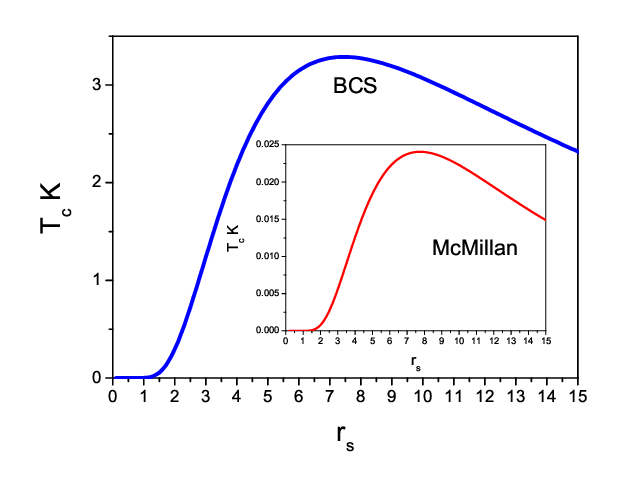}
\caption{Superconducting transition temperature of metallic hydrogen in
``jellium'' model calculated from the weak -- coupling BCS expression. 
At the insert: the same calculated from McMillan formula for intermediate 
coupling.}
\label{Tcjell}
\end{figure}

The upper boundary for the phonon frequency in this model is just given by the 
ion plasma frequency (we again consider the single -- charged ions):
\begin{equation}
\omega_0=\sqrt{\frac{4\pi ne^2}{M}}=2Ry\sqrt{\frac{3m}{M}}\frac{1}{r_s^{3/2}}
\label{plasm}
\end{equation}
where $n$ is conduction electron density. Note that the combination of
fundamental constants here is the same as in Eq. \ref{Tc_star}, but $r_s$ --
dependence is different.

Electron -- phonon and Coulomb dimensionless  coupling constants are given by 
\cite{VdM}: 
\begin{equation}
\lambda=\mu=\frac{1}{2\pi p_Fa_0/\hbar}\ln(1+\pi p_Fa_0/\hbar)=
\frac{r_s}{2\pi(\frac{9\pi}{4})^{1/3}}\ln\left[1+\pi\left(\frac{9\pi}{4}\right)^{1/3}r_s^{-1}\right]
\label{la_mu}
\end{equation}
Coulomb pseudopotential $\mu^{\star}$ is given by the usual expression:
\begin{equation}
\mu^{\star}=\frac{\mu}{1+\mu\ln\frac{E_F}{\omega_0}}
\label{mustar}
\end{equation}
where
\begin{equation}
\frac{E_F}{\omega_0}=\frac{1}{2}\frac{\left(\frac{9\pi}{4}\right)^{2/3}}{\sqrt{\frac{3m}{M}}}r_s^{-1/2}
\label{E_Fomeg}
\end{equation}

In principle the ``jellium'' model is more or less valid if we assume that
perturbation theory in Coulomb interaction can be applied, that is only for
$r_s<$1. However, we shall use all these expressions for rather wide region of 
$r_s$ as it was done in \cite{VdM}.

Now consider metallic hydrogen and take $M=m_p$. Then the characteristic 
phonon frequency $\omega_0$ given by Eq. \ref{plasm} is pretty large and changes
between circa 12000 K and 2000 K as $r_s$ changes between typical values of 1.0
and 3.5 (as in usual metals). As this frequency enters as a preexponential
factor in BCS -- like expressions for $T_c$ we can expect for it rather large
values \cite{Ginz,HTSC,Ash1}.

However, these expectations are not realized when we calculate coupling
constants.
Direct calculations show that $\lambda=\mu<1/2$ for all values of $r_s$ and
change between 0.17 -- 0.3 for most relevant values of $r_s=$1.0 -- 3.5,
signifying the weak coupling regime. For the same interval of $r_s$ the
Coulomb pseudopotential $\mu^{\star}$ changes between 0.1 and 0.15, as is 
usually assumed.

Then calculating $T_c$ for the wide range of $r_s$, using 
Equations \ref{plasm},\ref{la_mu},\ref{mustar},\ref{E_Fomeg} in the standard weak -- coupling BCS
expression ($\gamma=1.78$ is Euler constant):
\begin{equation}
T_c=\frac{2\gamma}{\pi}\omega_0\exp\left(-\frac{1}{\lambda-\mu^{\star}}\right)
\label{BCS}
\end{equation}
we obtain the results shown in Fig. \ref{Tcjell}. We can see the characteristic
maximum of $T_c$ at $r_s\sim$7, but the value of $T_c$ at this maximum only 
slightly exceeds 3K.

Even more pessimistic results are obtained if we calculate $T_c$ taking into
account intermediate coupling corrections and using McMillan formula \cite{VdM}:
\begin{equation}
T_c=\frac{\omega_0}{1.45}\exp\left[-\frac{1.04(1+\lambda)}{\lambda-\mu^{\star}(1+0.62\lambda)}\right]
\label{McM}
\end{equation}
which are shown at the insert in Fig. \ref{Tcjell}. Here we again observe the 
characteristic maximum of $T_c$, but it is less than 25 mK only! We must stress
the great sensitivity of these results to rather small change of the values of
main parameters, even such as the relatively weak $r_s$ -- dependence of the
Coulomb pseudopotential as given by Eq. \ref{mustar}.

However, we clearly see that in the weak (or intermediate) coupling approximations the
``jellium'' model can not produce high values of $T_c$ for metallic hydrogen
\cite{VdM}.

Somehow better results were obtained in Ref. \cite{VdM} by numerical solution of 
BCS integral gap equation with the explicit use of the screened 
``jellium'' -- like interaction kernel. Still, only the values of
$T_c$ not exceeding 30K were found at $r_s\sim 3$, which is also much lower than
the optimistic estimates of Refs. \cite{Ginz,HTSC,Ash1}. Actually the great 
difference obtained in Ref. \cite{VdM} between the results of direct solution
of the integral gap equation and more or less obvious use of the standard
BCS -- like of McMillan approximations is somehow obscure. We can only note,
that the authors of Ref. \cite{VdM} used the ``jellium''
model interaction as  direct input into BCS gap (integral) equation, though it 
is known since the famous KMK paper \cite{KMK}, that it should be actually 
replaced by some effective (smoothed) integral kernel. This remains to be done 
for the correct analysis and we just leave the use of KMK formalism in 
``jellium'' model for the future work.

As to real metallic hydrogen, we must stress that the ``jellium'' model is
only an oversimplified approximation, strictly valid for $r_s<1$ and neglecting
all the crystal structure effects in solid metallic hydrogen. At present there
is no consensus on the stable crystal structures of metallic hydrogen at all.
It is clear from our discussion above that the strong coupling regime can be
actually achieved, e.g. due to some lattice (phonon spectrum) softening 
effects, as was demonstrated in Refs. \cite{SavMaks,EGM}

\section{Conclusion}

Eliashberg theory remains the main theory, which completely explains the values
of the critical temperature in superconductors with electron -- phonon mechanism
of pairing. This theory is also applicable in the region of strong electron -- phonon
coupling, limited only by the applicability of adiabatic approximation, based on Migdal
theorem, which is valid in the vast majority of metals, including the new superhydrides
with record values of $T_c$. The values of (renormalized, physical) pairing coupling
constant $\lambda$ can surely exceed unity until the system possess the metallic
ground state. 

A number of simple expressions can be derived to estimate the maximal value
of $T_c$ in terms of experimentally measurable or calculable parameters like
characteristic (average) values of phonon frequencies and pairing coupling constant.
Actually this maximal value of $T_c$ is just determined by some ``game'' of
atomic constants. These estimates of the maximal possible $T_c$ are rather
optimistic, and the perspective for the experimental search for its higher
values, e.g. in hydrides, seems still very interesting.
However, all the present day data on superhydrides strongly indicate that all
these systems are actually very close to the strong coupling region of
Eliashberg theory, which indicate that the maximal values of $T_c$ for  metals
are probably more or less already achieved in these experiments.
In this respect the metallic hydrogen remains in our opinion probably the
most promising candidate system \cite{SavMaks,EGM}.

\medskip
\textbf{Acknowledgements} \par 

I am grateful to my late friends Daniel Khomskii and Eugene Maksimov
for many discussions related to the problems of high -- $T_c$ physics.

I am also grateful to Boris Altshuler and Andrei Chubukov for discussions
on specific heat instability, and to to Edward Kuchinskii for our recent 
discussions of the ``jellium'' model. 

I am remembering several meetings with Mikhail Eremets during different
conferences and seminars, which stimulated my interest in reanalysis of
superconductivity in electron -- phonon systems. 



\end{document}